\documentstyle[aps]{revtex}
\twocolumn
\begin{document}
\input{epsf}
\draft
\twocolumn[\hsize\textwidth\columnwidth\hsize\csname@twocolumnfalse\endcsname
\title{DYNAMICS OF SOLITONS AND QUASISOLITONS OF CUBIC THIRD-ORDER NONLINEAR 
SCHR\"{O}DINGER EQUATION}
\author{V.I. Karpman$^{1,}$ , 
J.J.Rasmussen$^{2}$  
and A.G. Shagalov$^{3}$ }
\address{$^{1}$Racah Institute of Physics, Hebrew University of Jerusalem, Jerusalem 91904, Israel
\\Dept. of Optics and Fluid Dynamics,  Risoe National Laboratory, Box 49, DK-4000 
Roskilde, Denmark,
\\ $^{3}$Institute of Metal Physics,
Ekaterinburg 620219, Russian Federation.} \maketitle
\begin{abstract}
The dynamics of soliton and quasisoliton solutions of cubic third order 
nonlinear Schr\"{o}dinger equation is studied. The regular solitons exist 
due to a balance between the nonlinear terms and (linear) third order dispersion; 
they are not important at small $\alpha_3$( $\alpha_3$ is the coefficient 
in the third derivative term) and vanish at $\alpha_3 \to 0$. The most essential, 
at small $\alpha_3$, is a quasisoliton 
emitting resonant radiation (resonantly radiating soliton). Its relationship 
with the other (steady) quasisoliton, called embedded soliton, is studied 
analytically and in numerical experiments. It is demonstrated that the 
resonantly radiating solitons emerge in the course of nonlinear evolution, 
which shows their physical significance.
\end{abstract}
\pacs{PACS numbers: 42.65.Tg, 52.35.Mw }
\vskip1pc] \narrowtext

{\bf I. INTRODUCTION}

\bigskip

Equations describing soliton processes are usually obtained by certain 
approximating procedures affecting nonlinearity and dispersion. Now, in 
extensive studies of ultrafast processes, the classical approximations often 
appear to be insufficient and higher order effects become of importance. A 
typical example is high speed systems like nonlinear transmission lines in 
femtosecond regime for soliton communications, etc. For such systems soliton 
solutions in classical sense do not generally exist. Only for very specific 
choices of parameters one can find localized solutions. Instead of regular 
solitons there may appear nonlocal steady or (and) unsteady soliton like structures 
(which may be called quasisolitons). These may, however, have 
significant importance for the nonlinear dynamics. 

In this paper we consider dynamics described by the extended third-order 
cubic nonlinear Schr\"{o}dinger (NLS) equation
$$
i\partial_T\Psi+\frac12\partial_X^2\Psi+|\Psi|^2\Psi+i\alpha_1|\Psi|^2\partial_X\Psi
+i\alpha_2\Psi\partial_X|\Psi|^2
$$
$$
+i\alpha_3\partial_X^3\Psi=0 \, ,
\eqno(1)
$$
which plays an important role in many nonlinear problems, in partiqular, in 
the nonlinear fiber optics [1,2]. The third derivative term describes higher 
order correction to the linear dispersion and the cubic terms with the first 
derivative are dispersive nonlinear corrections. Here, we assume that all
$\alpha_n$ ( $1 \le \alpha_n \le  3$) are real parameters. 

Eq.(1) has regular soliton solutions [3-5], vanishing at $|X|\to 0$. 
In particular cases, Eq.(1) may have exact {\it N} - soliton 
solutions [4] or even be integrable [5]. An important feature of the 
solutions describing regular solitons is that they are degenerating at
$\alpha_3 \to 0$ (see below), 
which substantially reduces their physical importance. 
Their numerical investigation at $\alpha_{1,2}\sim 1$
where done e.g. in Ref.[6] (see also references therein)

There also 
exist some solutions to Eq.(1), describing quasisolitons. One of them is a 
steady solution looking like a soliton type pulse embedded into a small 
amplitude plane wave (a soliton on plane wave pedestal). We call it {\it 
embedded} soliton (ES), using the word proposed in a different context in 
Ref.[7]. Apart from that, Eq. (1) has other type of quasisoliton solutions 
which describe soliton-like pulses permanently emitting resonantly generated 
radiation (see, e.g., Refs. [8-12] for the particular case at
$\alpha_1=\alpha_2=0$ and [13,14] 
for the full Eq.(1)). We will call them resonantly radiating solitons (RRS); 
they are unsteady because of losses caused by the radiation. The lifetime of 
radiating soliton is sufficiently large if 
$\alpha_3$ is small enough and, naturally, 
one can speak about the soliton only in this quasisteady case. (However, at 
long times, the losses caused by the radiation become essential for 
applications [2].) At $\alpha_3\to 0$, the RRS turns into 
the regular soliton of Eq.(1) 
without the third derivative. 

There is an interesting and important 
connection between the two types of quasisolitons, ES and RRS, and we 
demonstrate it in numerical simulations. We will also study the role of 
solitons and quasisolitons in the nonlinear processes described by Eq.(1). 
In partiqular, the regular solitons and RRS compete between themselves in 
nonlinear processes. As far as the regular solitons disappear 
at $\alpha_3\to 0$ and the RRS 
have short lifetimes at large $\alpha_3$, 
it is clear that at large $\alpha_3$ the regular 
solitons are more important while the RRS may play a decisive role at 
sufficiently small $\alpha_3$. The most interesting case is, 
of course, small $\alpha_3$ , 
because the third derivative term emerges as the result of an expansion. 
(The effect of next, 4th derivative, term can be seen in Refs.[13,14]; at 
certain relalionship between the coefficients before third and fourth 
derivatives there can be no radiation at all which, naturally, may happen 
also when other high-order terms are taken into account.) 

The paper is organized as follows. In Sec.I we describe important properties 
of Eq.(1), which are used below. In particular, we discuss regular soliton 
solutions [3], Galilean transformations and conservation laws [13,14]. The 
embedded solitons are studied in Sec.3. As we have already mentioned, the ES 
consists of a steady soliton-like pulse on the plane wave background. 
This structure is rater common in different nonlinear highly dispersive
systems [9-20].
At $\alpha_3\to 0$ , 
the plane wave disappears and the pulse turns into a regular soliton of Eq. 
(1) at $\alpha_3=0$ . The plane wave amplitude, increasing with $\alpha_3$ , 
may become unstable at sufficiently large $\alpha_3$ 
due to the modulational instability. Considering small 
$\alpha_3$, we show that the pulse part of ES is rather 
close to the pulse in RRS and 
the wavenumber of the plane wave coincides with the wavenumber of resonantly 
emitted radiation (by the RRS) . Next, we introduce the {\it cut off} 
operation, cutting off both wave wings from the ES (Sec.IV). Then we see 
that the remained soliton-like pulse is transformed into the RRS, emitting 
radiation only in {\it one direction} , according to the direction of the 
group velocity of resonant radiation. The resonant radiation disappears at 
$\alpha_1=6\alpha_3$. In this case, considered in Sec.V, we show that Eq.(1) can be 
transformed, by means of the Galilean transformation, to the complex 
modified Korteweg-de Vries (MKdV) equation . At particular initial 
conditions, it is reduced to the real MKdV equation which is integrable and 
therefore has {\it N} - soliton solutions [22]. This does not mean the complete 
integrability of Eq.(1) at $\alpha_1=6\alpha_3$ , 
because the reduction to the real MKdV equation 
is possible only for particular initial conditions. However, the Painlev\'{e} 
analysis [21] of the above mentioned {\it complex} MKdV equation shows 
that at $6\alpha_3= \alpha_1= 2 \alpha_2$ 
it possesses the Painlev\'{e} property. This is just the case when Eq. 
(1) is integrable [5]. If $\alpha_2=0$, we arrive at the case considered by Hirota [4] 
who has shown the existence of {\it N} - soliton solutions in this case. Our 
approach, however, is more general and (in fact, as in Ref. [23] ) we can 
see that the {\it N} -soliton solutions are possible even for nonvanishing 
$\alpha_2$ (at $\alpha_2\ne -3\alpha_3$ ). 
Then we present a numerical study of the initial value problem in the 
Hirota case:$\alpha_1=6\alpha_3$, $\alpha_2=0$  
taking as initial conditions complex pulses and one can see 
how the robust Hirota solitons are emerging from these pulses. In Sec.VI, we 
describe numerical experiments at $\alpha_1\ne 6\alpha_3$.
 Solving the initial value problem, we show that, at sufficiently 
small $\alpha_3$, the initial disturbances decay into the 
resonantly radiating solitons (RRS). On the other hand, at small $\alpha_3$, the 
Potasek- Tabor solitons [3] were not detected. This indicates that the RRS 
are important physical objects playing a significant role in nonlinear 
dynamics. In Sec. VII , a summary of obtained results is given. 
In Appendix A, the 
modulational instability of a plane wave, described by Eq.(1), is 
considered; it is helpful in the study of
the stability of embedded solitons.
In Appendix B we investigate analytically, by means of conservation laws, 
the evolution 
of RRS caused by the radiation. This analysis is in agreement with 
numerical results described in Secs.IV and VI.

\bigskip

{\bf II. IMPORTANT PROPERTIES OF EQ. (1)}

\bigskip

{\bf a. Exact soliton solutions}

\bigskip

First, we discuss the exact soliton solutions of Eq.(1). They can be written 
in the form 
$$
\Psi_s=a \; sech[b(X-V_sT)]\; e^{ikX-i\omega T}\;, 
\eqno(2a)
$$
$$
a^2=\frac{6\alpha_3}{\alpha_1+2\alpha_2}b^2 \;\;\; (\alpha_1 \ne -2\alpha_2), 
\eqno(2b) 
$$
$$
V_s=\kappa-3\alpha_3\kappa^2+\alpha_3b^2, 
\eqno(2c)
$$
$$
\omega=\frac12 \kappa^2-\alpha_3\kappa^3-\frac{1-6\alpha_3\kappa}{2}b^2, 
\eqno(2d)
$$
$$
\kappa=\frac{\alpha_1+2\alpha_2-6\alpha_3}{12\alpha_2\alpha_3}. 
\eqno(2e)
$$
In fact, these are the soliton solutions found by Potasek and Tabor [3] (with 
corrected misprints). We will call Eqs. (2a-e) Potasek-Tabor (PT) soliton 
solutions. At $\alpha_3=0$, PT solitons do not exist; this signifies that their 
existence is a result of a ballance between the nonlinear terms and the 
linear third order dispersive term . If $\alpha_2\ne 0$, from Eq. (2e) 
it follows that $\kappa$ is a 
fixed number, uniquely determined by the coefficients of Eq.(1). However, 
the solitons (2) exist even at $\alpha_2=0$ provided that
$$
\alpha_1=6\alpha_3. 
\eqno(3)
$$
In this case $\kappa$ can be {\it arbitrary} , because Eq. (2e) follows from the 
equation $12\alpha_2\alpha_3\kappa=\alpha_1-6\alpha_3+2\alpha_2$.

In case (3) and $\alpha_2=0$, Eq. (1) is the so called Hirota equation that 
can be transformed to the complex modified Korteweg-de Vries equation which has 
{\it N} -soliton solutions [4]. For the Hirota soliton, from Eq.(2b) it 
follows
$$
b=a. 
\eqno(4)
$$
In the other particular case, 
$6\alpha_3=\alpha_1=2\alpha_2$, Eq. (1) is integrable [5]. (See also Sec.V.) 
Some results for solitons (2) with $\alpha_2\ne 0$ 
were reported in Ref.[6] (see also 
references therein) . In particular, it was shown numerically that they 
emerge in a solution of initiall value problem. However, this result was 
obtained at $\alpha_1=\alpha_3=1$;
at small $\alpha_3$, as it will be shown below, they cannot compete with 
RRS. 

Apart from the solution (2a-e), describing "bright" solitons, which was 
called in Ref. [3] $sech$-family, Potasek and Tabor have also found 
$tanh$-family which 
will not be discussed here. (Some generalizations, with both 
$sech$ and $tanh$ terms, are 
derived in Ref. [27] by means of rather tedious algebra; the results of this 
paper are mostly contained in Ref.[3] and the generalizations [27], which are 
valid for very specific choices of the parameters, can be obtained by the 
much simpler approach of Potasek and Tabor.) The results of Ref.[3] are repeated 
also in some other papers (e.g.[23]).

\bigskip

{\bf b. Galilean transformation}

\bigskip

In the following we shall frequently use the Galilean transformation 
describing the transition to the reference frame moving with the velocity 
{\it V} . For Eq. (1) it reads 
$$
\Psi(X,T)=\psi(X-VT,T)\exp[i(KX-\Omega T)] \;, 
\eqno(5)
$$
where  $K$ and $\Omega$ are defined by equations
$$
V=K-3\alpha_3 K^2, 
\eqno(6)
$$
$$
\Omega=\frac12 K^2-\alpha_3 K^3. 
\eqno(7)
$$
The function $\psi(x,t)$ with 
$$
x=X-VT \;\;, \;\; t=T 
\eqno(8)
$$
satisfies the equation [13,14]
$$
i\partial_t\psi+\frac12 a_2\partial_x^2\psi+q|\psi|^2\psi+i\alpha_1|\psi|^2\partial_x\Psi
+i\alpha_2\psi\partial_x|\psi|^2
$$
$$
+i\alpha_3\partial_x^3\psi=0 \; ,
\eqno(9)
$$
with
$$
a_2=1-6\alpha_3 K, 
\eqno(10)
$$
$$
q=1-\alpha_1 K. 
\eqno(11)
$$

\bigskip

{\bf c. Conservation laws} 

\bigskip

One can check by means of straightforward calculations that Eq. (1) has the 
following integral of motion
$$
N=\int_{-\infty}^{\infty} |\Psi(X,T)|^2 dX \;. 
\eqno(12)
$$
At 
$$
\alpha_2=0, 
\eqno(13)
$$
we have two other conserving integrals [24, 13,14]
$$
P=\frac12 i \int_{-\infty}^{\infty} \left(
\Psi\partial_X\Psi^* - \Psi^*\partial_X\Psi
\right)dX \; , 
\eqno(14)
$$
$$
H=\int_{-\infty}^{\infty} \{
\frac12|\partial_X\Psi|^2 - \frac12|\Psi|^4 - \frac14 i\alpha_1 |\Psi|^2
(\Psi^*\partial_X\Psi - \Psi\partial_X\Psi^*) \\
$$
$$
-\frac12 i\alpha_3(\Psi^*\partial_X^3\Psi - \Psi\partial_X^3\Psi^*)
\}dX.
\eqno(15)
$$

\bigskip

{\bf III. EMBEDDED SOLITONS}

\bigskip

We start with the solutions to Eq. (1) of the form
$$
\Psi(X,T)=\chi(x)e^{i\Lambda t}, 
\eqno(16)
$$
where $x$ and $t$ are defined in (8). Substituting (16) into Eq.(1), we arrive at 
the ordinary differential equation
$$
-iV\partial_x\chi+\frac12\partial_x^2\chi+|\chi|^2\chi+i\alpha_1|\chi|^2\partial_x\chi
+i\alpha_2\chi\partial_x|\chi|^2
$$
$$
+i\alpha_3\partial_x^3\chi=\Lambda\chi \; .
\eqno(17)
$$
Imposing periodic boundary conditions at the ends of a sufficiently broad 
interval and considering {\it V} and $\chi(0)$ as given parameters and 
$\Lambda$ as the 
eigenvalue, we arrive at a nonlinear eigenvalue problem which can be 
numerically solved by a kind of shooting method. An example of such solution 
at $\alpha_2=0$ and
$$
\alpha_1=-0.3\;,\;\;
\alpha_3=-0.1\;,\;\;
\chi(0)=1.5\;,\;\;
V=0 
\eqno(18)
$$
is shown in Fig.1(a,b) . It is a nonlocal steady pulse with 
$|\Psi(x)|_{max}=|\Psi(x_0)|$ and small 
symmetric "wings". At large $|x-x_0|$ we can linearize Eq. (17); then we see that 
$$
\chi(x)\approx {\text{const}}\; e^{i\kappa x}  \;\;\;\;\;
(|x-x_0|\gg 1) \;,
\eqno(19) 
$$
where $\kappa$ is a real root of cubic algebraic equation
$$
\Lambda-\kappa V+(1/2)\kappa^2-\alpha_3\kappa^3=0 \;. 
\eqno(20)
$$ 
Finding $\Lambda$ from the numerical solution of the eigenvalue
poblem and $\kappa$ from (20), we have for the case (18)
$$
\Lambda=0.979 \;,\;\;\;\; \kappa=-5.34 \;. 
\eqno(21)
$$
On the other hand we can determine $\kappa$ directly from 
$\chi(x)$  which is found numerically 
from Eq. (17), together with $\Lambda$. From Fig. 1b we see that the numerically 
found $\chi(x)$ has indeed asymptotic behavior (19) with $\kappa\approx -5.3$ 
which is very close to the 
root of Eq. (20), written in (21). This agreement is an evidence of the 
correctness of the numerical solution of eigenvalue problem by the shooting 
method.
In a similar way, the solution of eigenvalue problem (17) and Eq. (20) for 
$\alpha_2=0$ and
$$
\alpha_1=-0.8\;,\;\;
\alpha_3=-0.1\;,\;\;
\chi(0)=1.5\;,\;\;
V=-0.225 
\eqno(22a) 
$$
gives
$$
\Lambda=1.209 \;,\;\;\;\; \kappa=-5.03 \;. 
\eqno(22b)
$$
\begin{figure}[h] \vspace{-.5cm} \hspace{0.0cm}
\vspace*{3cm}
\caption{ Numerical solution of Eq. (17) at parameters (18). 
(a) $\chi$ versus $x$; solid line: $|\chi(x)|=|\Psi(x)|$; 
dashed line: $Re\chi(x)$. (b) $arg\chi$ versus $x$; at 
sufficiently large $|x-x_0|$, $arg\chi(x)$ is a linear 
function: $arg\chi\approx \kappa |x-x_0|$, which permits to 
measure $\kappa$ ($\kappa\approx -5.2$).}
\end{figure}
The asymptotic behavior of numerically found $\chi(x)$ gives the same 
$\kappa\approx -5.03$. 

For the other case with $\alpha_2=0$ and
$$
\alpha_1=-1.1\;,\;\;
\alpha_3=-0.1\;,\;\;
\chi(0)=1.5\;,\;\;
V=0.35 
\eqno(23a)
$$
we have
$$
\Lambda=1.937 \;,\;\;\;\; \kappa=-6.12 \;. 
\eqno(23b)
$$
while from the numerically found $\chi(x)$ we obtain $\kappa=-6.1$. 

In fact, similar quasisoliton solutions, looking like embedded solitons, 
were obtained, by different approaches, for other highly dispersive systems 
as well [15-17,9,11,12,17-20].

Now consider the embedded solitons, starting from Galilean transformation 
(5). Writing it in the form
$$
\Psi(X,T)=\psi(x,t)\; e^{iKx}\; {\text{exp}}[i(KV-\Omega) t]
\eqno(24)
$$
and assuming that 
$$
\psi(x,t)=\tilde{\psi}(x){\text{exp}}\left(i\frac12\lambda^2 t\right), 
\eqno(25) 
$$
where $\lambda$ is a constant parameter, we compare (24) with (16). Then
we have
$$
\tilde{\psi}(x)=\chi(x) {\text{exp}}(-iKx)\;, 
\eqno(26)
$$
$$
\lambda^2=2(\Lambda+\Omega-KV)\;. 
\eqno(27)
$$

From (25) and (9) we arrive at the following equation for 
$\tilde{\psi(x)}$
$$
-\frac12 \lambda^2\tilde{\psi}+\frac12 a_2\partial_x^2\tilde{\psi}+
q|\tilde{\psi}|^2\tilde{\psi}+i\alpha_1|\tilde{\psi}|^2\partial_x\tilde{\psi}
+i\alpha_2\tilde{\psi}\partial_x|\tilde{\psi}|^2
$$
$$
+i\alpha_3\partial_x^3\tilde{\psi}=0 \, ,
\eqno(28)
$$
Substituting here (26), (27) and taking into account (6), (7) and (10), (11) 
we have Eq.(17) as one should expect. From Eq. (26) it is seen that
$|\tilde{\psi}|^2\ll 1$ at $|x|\gg 1$. 
Linearizing Eq. (28) we obtain 
$$
\tilde{\psi}\sim e^{ikx} \;\;\;\;\; (|x|\gg 1), 
\eqno(29) 
$$
where $k$ is a root of the equation
$$
2\alpha_3 k^3-a_2k^2-\lambda^2=0 \;. 
\eqno(30)
$$
Substituting (19) and (29) into Eq. (26), we see that
$$
\kappa=k+K. 
\eqno(31)
$$
Then, using (6), (7) and (10), (27) we can easily prove that Eqs. (30) and 
(20) are equivalent, as it should be.

At small $\alpha_3$, the solution of Eq.(28) can be written as
$$
\tilde{\psi}(x)=[u_s(x)+f(x)]\text{exp} [i\phi_s(x)]. 
\eqno(32)
$$
with small $f(x)$. Here $u_s(x)$ and $\phi_s(x)$ are defined by the requirement that
$$
F(x)=u_s(x)\text{exp} [i\phi_s(x)]
\eqno(33)
$$
is a soliton solution of Eq. (28) without the last term i.e.
$$
-\frac12 \lambda^2 F+\frac12 a_2\partial_x^2 F+
q|F|^2 F+i\alpha_1|F|^2\partial_x F
+i\alpha_2 F\partial_x|F|^2=0
\eqno(34)
$$
and $F(x)\to 0$ at $x=\pm \infty$. Solving Eq. (34), we have 
$$
u_s(x)=\lambda\sqrt{\frac{2p}{q}} \left[\text{cosh}
\left( \frac{2\lambda}{\sqrt{a_2}}x\right)+p\right]^{-1/2} \;, 
\eqno(35)
$$
$$
\phi_s(x)=-\frac{\alpha_1+2\alpha_2}{2A} \text{arctan}
\left[\sqrt{\frac{1-p}{1+p}}\text{tanh}\left(\frac{\lambda x}{\sqrt{a_2}}\right) \right]
\;, 
\eqno(36)
$$
$$
p=\frac{\sqrt{a_2}q}{\sqrt{4A^2\lambda^2+a_2q^2}}, 
\eqno(37)
$$
$$
A^2=\frac{4\alpha_1(\alpha_1+2\alpha_2)-(\alpha_1+2\alpha_2)^2}
{12}  \;. 
\eqno(38)
$$
Eqs. (35)-(38) were obtained in Ref. [14] for a pulse part of RRS. From (35) 
it follows that the soliton amplitude is
$$
u_0=\lambda\sqrt{\frac{2p}{(1+p)q}}
\eqno(39)
$$
and its width is given by
$$
\delta=\sqrt{a_2}/\lambda \;. 
\eqno(40)
$$
Thus one must require $a_2>0$ and from Eq. (10) it follows that at
 $\alpha_3 K>0$,
$$
|K|<\frac16 |\alpha_3|   \;, 
\eqno(41)
$$
which is a restriction on the soliton velocity (6).

A small term $f(x)$ in Eq.(32) expresses the effect of the third order dispersion. 
At $|x|\sim \sqrt(a_2)/\lambda$, or 
less, it describes a modification of the pulse, arising due to the last term 
in Eq. (28), while at large $x$
$$
f(x)\approx \tilde{\psi}(x)\sim e^{ikx}
\eqno(42)
$$
where $k$ is a root of Eq. (30). Note, that this equation was derived in Ref. 
[14] for the wavenumber of the resonantly generated radiation by the RRS in 
the reference frame where the RRS is at rest. From all that we conclude that 
(32) coincides with the asymptotic expression for the RRS and its radiation 
at large {\it t}.

Let us now compare the embedded solitons, obtained above numerically, with 
the solution (32) at condition (13) . From Eq.(38) we have
$$
A=\frac12 |\alpha_1| \;. 
\eqno(43)
$$
Eqs. (6) and (10) give
$$
K=\frac{1-\sqrt{1-12\alpha_3 V}}{6\alpha_3} \;, 
\eqno(44)
$$
$$
a_2=\sqrt{1-12\alpha_3 V}  \;. 
\eqno(45)
$$
Using (39) and (37), we obtain 
$$
u_0^2=\frac{\sqrt{a_2}}{2A^2}(\sqrt{4A^2\lambda^2+q^2 a_2} - 
q\sqrt{a_2}) \;, 
\eqno(46)
$$
where {\it q} is defined in (11). From (27) and (6),(7) we find
$$
\lambda^2=2\left(\Lambda-\frac12 K^2+2\alpha_3 K^3 \right). 
\eqno(47) 
$$
Then we have for the case (18), (21), which is shown in Fig. 1:
$u_0\approx 1.37$. For the case (22) $u_0\approx 1.41$, and for (23) 
$u_0\approx 1.42$. So, in all three cases we approximately have
$u_0\approx 1.4$ which is rather close to $\chi(0)=1.5$, 
assumed in the nonlinear eigenvalue problem for all 
three cases. This supports the conclusion that expressions (35) and (36) 
approximately describe the pulse in the embedded soliton 
(at small  $\alpha_3$ ).

Now consider $f(x)$. At small $\alpha_3$, the roots of Eq. (30) have 
simple analytical expressions [14]. Neglecting the first term in (30), 
we have two smallest roots
$$
k\approx \pm i \frac{\lambda}{\sqrt{a_2}}    \;. 
\eqno(48)
$$
Substituting this into (42), we have
$$
\tilde{\psi}\approx f(x) \sim \text{exp}(\mp \lambda/\sqrt{a_2} x)   \;. 
\eqno(49)
$$
This is in agreement with the asymptotic behavior of expression (33) at 
large $x$. The third root can be approximately obtained if one neglects the 
last term in (30). This gives
$$
k\approx a_2/2\alpha_3  \;, 
\eqno(50)
$$
which approximately determines the wavenumber of the plane wave in the 
wings. Expressions (50) and (31) approximately give the roots of Eq. (20) in 
analytical form. From (50) it follows
$$
\text{sgn}\; k=\text{sgn} \;\alpha_3          \;. 
\eqno(51)
$$
And, finally, from the results of Ref. [14] it follows that at small 
$\alpha_3$
$$
f(x)\approx B\left(\frac{\sqrt{a_2}|k|}{A}\right)^{\frac12}
\text{exp}\left[-\frac{\pi\sqrt{a_2}|k|}{4\lambda}\left(1+\frac{2}{\pi}
\text{arcsin} p \right)\right] e^{ikx}
\; , 
\eqno(52)
$$
where {\it B} is a complex constant with 
$|B|\sim 1-10$. This expression is valid at
$$
\frac{\sqrt{a_2}|k|}{\lambda} \gg 1 \;, 
\eqno(53)
$$
i.e. when the wavenumber $k$ is much larger than the inverse width of the 
soliton (33). In expression (52), it is also assumed that 
$2A\lambda\gg\lambda/\sqrt{a_2}|k_{1,2}|$
(this does not exclude $A\lambda\ll 1$). If
$$
2A\lambda\sim \lambda/\sqrt{A_2}|k_{1,2}|
\;\;\;\; (\text{or}\;\; 2A\lambda<\lambda/\sqrt{a_2}|k_{1,2}| ) , 
\eqno(54)
$$
which may be satisfied only at $p\approx 1$ and $q\approx 1$, we have 
$$
f(x)=B\sqrt{a_2}|k|\text{exp}\left(-\frac{\pi\sqrt{a_2}|k|}{2\lambda} \right) 
e^{ikx} \;. 
\eqno(55)
$$

It is easy to check that we can arrive at Eq. (55) by substituting in Eq. 
(52) the first of conditions (54) and {\it p=1} . 

From the all foregoing, one can see a connection between the embedded and 
resonantly radiating solitons. In the next section we present numerical 
experiments disclosing this connection from another side.

\bigskip

{\bf IV. THE CUT OFF OPERATION}

\bigskip

Let us define the cut off operation transforming the embedded soliton into 
an isolated pulse. Turning to the function
$\chi(x)$ in (16), we write
$$
\chi_{cut}(x)=\chi(x) r(x)           \;, 
\eqno(56)
$$
where $r(x)$ is a cutting factor which we take in the form
$$
r(x)=\frac12\left[\text{tanh}\left(\frac{x-x_0+\Delta x}{\gamma}\right)     
-\text{tanh}\left(\frac{x-x_0-\Delta x}{\gamma}\right)\right]        \;. 
\eqno(57)
$$
Here $x_0$ is the center of the pulse and $\Delta x>0$ is the width 
of the cutted function $\chi_{cut}(x)$. 
According to (57), $r(x)$ vanishes at $|x-x_0|\to\infty$ and the positive 
parameter $\gamma$ characterizes the 
"sharpness" of vanishing. Assuming that $\gamma$ is small enough 
and $\Delta x$ is such that the factor $r(x)$
cuts off only the wings without essential disturbing the pulse in 
$\chi(x)$ ( $\gamma$ and $\Delta x$ can be properly chosen in numerical tests)
 we then take the pulse $\chi_{cut}(x)$ as the initial condition to Eq.(1) 
(see Fig.2a)
$$
\Psi(X,0)=\chi_{cut}(X)      \;. 
\eqno(58) 
$$
In Fig.2b one can see that the cutted pulse emits radiation at 
$t>0$. However, on the left hand side of the pulse the radiation 
spreads out with time; this shows that on the left hand side there is a 
transient radiation, emitted at small {\it t} due to initial condition (58). 
On the contrary, on the right hand side we see a wave train with 
approximately constant amplitude, with the front propagating to the right; 
so, the length of the wavetrain increases with time. Therefore the cutted 
pulse permanently emits radiation to the right. The spectrum, shown in 
Fig.1c, has a peak at $\kappa$ approximately equal to the root of Eq. (20) with 
account of finite velocity of the pulse at $t=138$ (it is still rather close 
to $\kappa$ from 
Eq. (21), i.e. to the wavenumber of the wing waves in embedded soliton).

Analyzing the time behavior of $\text{arg}\Psi$, we find that
$d(\text{arg}\Psi)/dt=\Lambda(t)$
is a slow function of {\it t}, with $\Lambda(0)\approx 0.98$ and
$\Lambda(130)\approx 0.85$. Note that this $\Lambda(0)$ coincides, with a 
good accuracy, with (21). Similar results were obtained for cutted pulses in 
cases (22) and (23). 

We conclude that the cutted pulses are radiating solitons; and the radiation 
is permanently emitted only in one direction. As far as $\kappa$ is connected 
with $k$
from Eq. (31), which is the real root of Eq. (30), obtained from the 
resonant condition, we conclude that the cut off operation transforms ES ino 
RRS. 
The front of radiated wavetrain propagates with the group velocity 
$U(k)$ given by [14]
$$
U(k)=a_2 k-3\alpha_3k^2 \approx -a_2/4\alpha_3  \;. 
\eqno(59)
$$
\begin{figure}[h] \vspace{-.5cm} \hspace{0.0cm}
\vspace*{3cm}
\caption{Numerical solution of Eq. (1) at initial condition (58) and 
$\alpha_1=-0.3$, $\alpha_2=0$, $\alpha_3=-0.1$.
(a) The initial cutted pulse. Full line: $|\Psi|$ versus {\it x} at{\it T=0}; 
dotted line depicts the corresponding embedded soliton. 
(b) The cutted pulse at {\it T=138} ; full line: $|\Psi|$ versus {\it x}; 
dashed line: $Re\Psi$ versus {\it x} . (c)$log_{10}|\Psi_k|$ 
versus $\kappa$( $\Psi_k$ is the Fourier transform of $\Psi(x)$; 
one can see a peak near the root of resonance equation (20).}
\end{figure}
From this it follows that
$$
\text{sgn}U(k)=-\text{sgn}k=-\text{sgn}\alpha_3        \;. 
\eqno(60) 
$$
Therefore at $\alpha_3<0$ it should be $k<0$ and $U(k)>0$. 
This means that the soliton in Fig. 2a 
should permanently emit radiation to the right
while the peak in the spectrum of the wavetrain should be at negative $k$. 
This is in agreement with the results presented in Figs. 2b and 2c. 

To describe analytically the whole system after the cut off, one can use at 
small $\alpha_3$ the equation [14] 
$$
\tilde{\psi}(x,t)=[u_s(x)+\eta(x,t)]\text{exp}[i\phi_s(x)]            \;, 
\eqno(61)
$$
where $u_s(x)$ and $\phi_s(x)$ are given (in adiabatic approximation) 
by Eqs. (35) and (36) and $\eta(x,y)$, 
at large $x$ and $t$, has the following asymptotic expression 
$$
\eta(x,t)\approx f(x)\Theta(Ux)\Theta(|U|t-|x|)         \;. 
\eqno(62)
$$
Here $f(x)$ is given by Eq. (52) and $\Theta(Y)$ is the Heaviside function
$$
\Theta(Y)=1  \;\;(Y>0) \;, 
\Theta(Y)=0  \;\;(Y<0) \;, 
\;. 
\eqno(63)
$$
Eq. (62) expresses that the soliton radiates in the direction of group 
velocity $U$ and the radiation front propagates with the velocity $|U|$. The 
mentioned above adiabatic approximation means that at sufficiently small
 $\alpha_3$, 
the radiation is so small that the soliton parameters in
$u_s(x)$ and $\psi_s(x)$
(as well as the wavenumber $k$) can be considered as constant. However, the 
soliton losses may be essential at large times. 
\begin{figure}[h] \vspace{-.5cm} \hspace{0.0cm}
\vspace*{3cm}
\caption{Temporal behavior of the radiating soliton at 
$\alpha_1=-0.3$, $\alpha_2=0$, $\alpha_3=-0.1$. (a) $|\Psi|_{max}$ versus {\it 
T=t} ; (b) the soliton position versus {\it t}.}
\end{figure}

The variation of soliton parameters, caused by the radiation, can be 
estimated by means of the integrals of motion (Appendix B). A 
vast information about the soliton evolution, caused by the radiation, 
follows from the numerical solution of Eq.(1). For the parameters used in 
Figs. 2, the results following from this solution are presented in Fig.3. 
One finds that the mean value of $|\Psi|_{max}=u_0(t)$, 
shown in Fig.3a, decreases 
logarithmically, similar to what was found for the case 
$\alpha_1=\alpha_2=0$ [25,26]. Analyzing 
the soliton position (Fig.3b) one can find that the soliton velocity $V(t)$
increases from $V(0)=0$ to $V(138)\approx 0.29$, i.e. the soliton is 
{\it accelerating} and the function $V(t)$ 
increases (also logarithmically). Now, using $\Lambda(0)\approx 0.98$, 
$\Lambda(138)\approx 0.85$ and Eqs. (47),(44), we have 
$\lambda(0)\approx 1.4$, $\lambda(138)\approx 1.28$. 
Therefore, the soliton width $\Delta=\sqrt{a_2}/\lambda$ increases from 
$\Delta(0)\approx 0.71$ to $\Delta(138)\approx 0.99$, i.e. the radiating 
soliton is widening with time . All that can be considered as a numerical 
confirmation of analytical results obtained in Appenix B from the 
conservation laws, in particular that
$$
\text{sgn}\frac{dV}{dt}=-\text{sgn}k=\text{sgn}U
\;, 
\eqno(64)
$$
which means that the soliton is accelerating in the direction of the group 
velocity of resonant radiation.

\bigskip

{\bf V. SPECIAL CASE $\alpha_1=6\alpha_3$. HIROTA SOLITONS}

\bigskip

The coefficient {\it q} in Eq. (9) disappears at
$$
K=1/\alpha_1 \;. 
\eqno(65)
$$
Substituting this in Eq. (10) , we have 
$a_2=1-6\alpha_3/\alpha_1$. Therefore, at condition (3) the 
coefficient $a_2$ also vanishes and Eq. (9) takes the form
$$
\partial_t\psi+6\alpha_3|\psi|^2\partial_x \psi+\alpha_2\psi\partial_x|\psi|^2
+\alpha_3\partial_x^3\psi=0
\;. 
\eqno(66)
$$
\noindent
which is related to Eq.(1) by a Galilean transformation (5),(8) with
$$
V=\frac{1}{12\alpha_3}\;,
\;\;K=\frac{1}{6\alpha_3} \;,
\;\;\Omega=\frac{1}{108\alpha_3^2} . 
\eqno(67)
$$
Consider a particular solution of Eq.(66)
$$
\psi(x,t)=e^{i\theta}\zeta(x,t)
\;, 
\eqno(68)
$$
where $\theta=\text{const}$ and $\theta$, $\zeta(x,t)$ are real.
Then Eq. (66) is reduced to the modified Korteweg-de Vries (MKdV) equation
$$
\partial_t\zeta+2(3\alpha_3+\alpha_2)\zeta^2\partial_x\zeta+
\alpha_3\partial_x^3\zeta=0
\;. 
\eqno(69)
$$
It is completely integrable and, in particular, has exact {\it N} - 
soliton solutions [22] if $\alpha_2\ne -3\alpha_3=-(1/2)\alpha_1$ 
[cf. with Eq. (2b)]. As far as 
Eq. (66) is reduced to Eq. (69) only for particular initial conditions, it 
may not be , generally, integrable; however, one can assert that it also 
possess {\it N} - soliton solutions at any $\alpha_2\ne -(1/2)\alpha_1$, 
which is a generalization of 
the Hirota solution at $\alpha_2=0$. Below, we call Eq. (66) complex MKdV equation. 
The Painlev\'{e} test [21], applied to Eq. (66), shows that it
has the Painlev\'{e} property if $\alpha_2=3\alpha_3$, i.e.
$$
6\alpha_3=\alpha_1=2\alpha_2
\;. 
\eqno(70) 
$$
This is just the integrability condition for Eq.(1) found by Sasa and 
Satsuma [5]. These conclusions are in agreement with those following
from the Painlev\'e test of Eq.(1) [28,23].

If, in addition to condition (3), $\alpha_2=0$ (Hirota case [4]), we arrive 
at the complex MKdV equation
$$
\partial_t\psi+6\alpha_3|\psi|^2\partial_x \psi
+\alpha_3\partial_x^3\psi=0
\;. 
\eqno(71)
$$
The soliton solution to this equation is 
$$
\psi_s(x,t)=a\; \text{sech}[a(x-ct)]\text{exp}[i(px-\sigma t)]
\eqno(72)
$$
with arbitrary {\it p} and
$$
c=-3\alpha_3p^2+\alpha_3a^2\;,
\;\;\; \sigma=-\alpha_3 p^3+3\alpha_3 p a^2 \; . 
\eqno(73) 
$$
Then
$$
\Psi_s(X,T)=a\; \text{sech}[a(X-V_sT)]\text{exp}[i(\kappa X-\omega T)]
\;, 
\eqno(74) 
$$
where
$$
V_s=V+c\;,
\;\;\; \kappa=K+p \;,
\;\;\; \omega=\sigma+\Omega+pV \;. (75)
$$
Therefore $\kappa$ is arbitrary and $V_s(\kappa)$, $\omega(\kappa)$ 
coincide with Eqs. (2c,d) respectively. This 
means that Eqs. (74) and (75) indeed describe the Hirota solitons, mentioned 
in Sec.2a. It is also seen that though the Hirota conditions do not ensure 
the integrability ( Painlev\'{e} tests are negative !), Eq. (1) in Hirota case 
definitely has exact {\it N} -soluton solutions. 

Let us now consider numerical solutions of the initial value problem to 
Eq.(1) in Hirota case 
$$
\alpha_1=6\alpha_3=-0.6\; ,\;\;\; \alpha_2=0 . 
\eqno(76)
$$
As initial condition, we choose a complex pulse
$$
\Psi(X,0)=A\text{sech}\left[\frac12(X-X_0)\right]
\text{exp}[iC(X-X_0)]
\eqno(77)
$$
with {\it A=1 , C=0.5} and $X_0=85.3$. The structures of 
$\Psi(X,T)$ at {\it T=0} and {\it T=168} 
are shown in Figs. 4(a,b). We see that $\Psi(X,T)$ evolves into two pulses, 
well separated at {\it T=168} . The spetrum of $\Psi(X,T)$ at {\it T=168} 
(Fig.5) gives an 
evidence that, in addition to these pulses, no significant wavetrains are 
produced at initial condition (77). The time variation of the amplitude of 
the pulse with largest amplitude, $|\Psi^{(1)}|$, is shown in Fig.6. 
It approaches the 
constant limit, approximately equal to 1.5. The velocity of that pulse 
approaches the limit approximately equal to 0.35 . These results suggest 
that the leading pulse asymptotically approaches a regular soliton with 
$a^{(1)}\approx 1.5$ and $V_s^{(1)}\approx 0.35$.
 The shapes of $\Psi(X,T)$, obtained numerically for $T>150$, confirm this conclusion. Then, 
using Eqs. (2c,d) and (4), we can calculate the shifted frequency 
$\omega'=\omega(k)-kV_s^{(1)}$, which is 
the soliton frequency in the reference frame where the soliton is at rest. 
The plot of  $\omega'$ versus $V_s$, 
shown in Fig.7, gives $\omega'(0.35)\approx 1.5$. This must be equal to 
$\text{arg}\Psi^{(1)}$. On the other hand, the numerical solution permits 
to obtain $\text{arg}\Psi^{(1)}$
versus time in the frame where the pulse is at rest. This also gives 
$\text{arg}\Psi^{(1)}\approx 1.5$, at 
sufficiently large {\it T} . Similar results were obtained for the second 
pulse. Thus we conclude that the initial {\it complex} pulse (77) decays 
into two Hirota solitons. Qualitatively similar results were obtained for 
the initial pulse
\begin{figure}[h] \vspace{-.5cm} \hspace{0.0cm}
\vspace*{3cm}
\caption{ $\Psi$ versus $x$ at $\alpha_1=-0.6$, $\alpha_2=0$, $\alpha_3=-0.1$ 
and initial condition (77); solid line: $|\Psi|$, dashed line: $Re \Psi$ 
(a){\it T=0} . (b) {\it T=168}.}
\end{figure}
$$
\Psi(X,0)=\text{exp}\left[-\frac{(X-X_0)^2}{16}+
i\frac{X-X_0}{2}   \right]\;, \;\;\;
X_0=85.3 \;. 
\eqno(78)
$$
\begin{figure}[h] \vspace{-.5cm} \hspace{0.0cm}
\vspace*{3cm}
\caption{ $log_{10}|\Psi_{\kappa}|$ versus $\kappa$ at 
$\alpha_1=-0.6$, $\alpha_2=0$, $\alpha_3=-0.1$ and {\it T=168}.}
\end{figure}
\begin{figure}[h] \vspace{-.5cm} \hspace{0.0cm}
\vspace*{3cm}
\caption{Temporal behavior of the largest soliton amplitude at
$\alpha_1=-0.6$, $\alpha_2=0$, $\alpha_3=-0.1$ and initial 
condition (77).}
\end{figure}
Other features of the dynamics of Hirota solitons, in particular their 
collisions , will be examined in another paper.

\bigskip

{\bf VI. PULSE EVOLUTION AT $\alpha_1\ne 6\alpha_3$, $\alpha_2=0$}

\bigskip

Here, we report on the numerical solutions of Eq.(1) in general case. First 
we take, as initial condition, the pulse (77) with {\it A=1.9, C=0} and 
$X_0=100$ and assume that 
$$
\alpha_1=-0.5\;, \;\;\; \alpha_3=0.1\;.
\eqno(79)
$$
The behavior of solution at these parameters is shown in Fig.8, where the 
case $\alpha_3=0$ (with $\alpha_1=-0.5$) is also presented for comparison. 
In the latter case, the 
initial pulse splits into four solitons of the form (33), (35)-(38), with $a_2=1$
and moving to the left (Fig.8a). On the other hand, in the case (79) the 
initial pulse splits into three {\it radiating} solitons. Two of them 
propagate to the right and the smallest one - to the left. 
\begin{figure}[h] \vspace{-.5cm} \hspace{0.0cm}
\vspace*{3cm}
\caption{$\omega'=\omega-\kappa V_s$ versus $V_s$ for the largest soliton.}
\end{figure}
\begin{figure}[h] \vspace{-.5cm} \hspace{0.0cm}
\vspace*{3cm}
\caption{Development of the initial pulse (77) with {\it A=1.9, C=0, $X_0=100$} 
(dashed line). 
(a) $\alpha_1=-0.5$, $\alpha_2=0$, $\alpha_3=0$ {\it t=48}, 
(b) $\alpha_1=-0.5$, $\alpha_2=0$, $\alpha_3=-0.1$ {\it t=48}, 
(c) $\alpha_1=-0.5$, $\alpha_2=0$, $\alpha_3=-0.1$ {\it t=175}.}
\end{figure}
\begin{figure}[h] \vspace{-.5cm} \hspace{0.0cm}
\vspace*{3cm}
\caption{Plots of $X_0(t)$ for three largest solitons shown in Fig.8.
 Dashed line: $\alpha_1=-0.5$, $\alpha_2=0$, $\alpha_3=0$;
solid line: $\alpha_1=-0.5$, $\alpha_2=0$, $\alpha_3=-0.1$}
\end{figure}
\begin{figure}[h] \vspace{-.5cm} \hspace{0.0cm}
\vspace*{3cm}
\caption{Fourier spectra of the solutions shown in Fig.8;
(a) $\alpha_1=-0.5$, $\alpha_2=0$, $\alpha_3=0$; {\it t=48}
(b) $\alpha_1=-0.5$, $\alpha_2=0$, $\alpha_3=-0.1$; {\it t=175}.}
\end{figure}
The radiating 
wavetrains propagate to the right with the group velocities larger than the 
soliton velocities (Figs. 8b and 8c); this is in agreement with Eq. (60). 
The time dependence of the soliton coordinates is shown in Fig.9 . 
At $\alpha_3$, the 
soliton velocities are constant. At $\alpha_3=-0.1$, the solitons are 
accelerating in {\it positive} direction, which agrees with Eq. (64).

The Fourier spectra for both cases, 
$\alpha_3=0$ and $\alpha_3=-0.1$, are shown in Fig. 10. The spectrum 
at $\alpha_3=0$ exhibits no resonant radiation (Fig.10a) while at 
$\alpha_3=-0.1$ we see three distinct 
resonant peaks with negative $\kappa$, which are the wave numbers of the resonant 
radiation emitted by the solitons in the "laboratory" frame (Fig.10b). The 
difference between the wave numbers follows from Eq. (20) which is valid, as 
we have seen, both for embedded and radiating solitons. As far as the 
solitons have different velocities and amplitudes (the latters are 
determined by the soliton parameter $\Lambda$), they have different wavenumbers 
because $\kappa$, being a root of Eq. (20), depends on the soliton velocity and 
amplitude.
\begin{figure}[h] \vspace{-.5cm} \hspace{0.0cm}
\vspace*{3cm}
\caption{The amplitude of the first soliton versus time. 
(a) $\alpha_3=0$, (b) $\alpha_3=-0.1$.}
\end{figure}
\begin{figure}[h] \vspace{-.5cm} \hspace{0.0cm}
\vspace*{3cm}
\caption{The amplitude of the second soliton versus time. 
(a) $\alpha_3=0$, (b) $\alpha_3=-0.1$.}
\end{figure}
An important difference between the two cases, $\alpha_3=0$ and 
$\alpha_3\ne 0$, is seen in Figs. 11 and 
12. In the first case, the average soliton amplitudes have constant limits 
at $t\to \infty$ while in the second case they must slowly decrease because of the soliton 
radiation.

In the aforegoing, we studied the case when $\alpha_1\alpha_3\ge 0$. 
Now let us consider $\alpha_1\alpha_3 \le 0$. We take
$$
\alpha_1=-0.6\;,
\;\;\; \alpha_3=0.1 \;, 
\eqno(80)
$$
\begin{figure}[h] \vspace{-.5cm} \hspace{0.0cm}
\vspace*{3cm}
\caption{The RRS at parameters $\alpha_1=-0.6, \alpha_3=0.1$. 
Full line: $|\Psi|$, dotted line : $Re\Psi$.}
\end{figure}
\begin{figure}[h] \vspace{-.5cm} \hspace{0.0cm}
\vspace*{3cm}
\caption{Spectral distribution at $\alpha_1=-0.6, \alpha_3=0.1$.}
\end{figure}
and initial condition (77) with $C=1/2$, $X_0=850$. 
The solution at $T=370$ is shown in Fig. 13. We 
see a pulse with radiation on both its sides. An analysis shows that on the 
left hand side there is the resonantly generated radiation (in agreement 
with Eq. (60), where now $\alpha_3>0$, $k>0$, $U<0$). 
On the right hand side, there is a 
radiation composed from the harmonics of continuous spectrum; it satisfies 
the linearized Eq.(1) and has been emitted in the transient period of time. 
The spectal distribution is shown in Fig.14. One can see two spectral 
maximums at positive wavenumbers. The narrow one corresponds to the resonant 
radiation, while the broader peak is composed of the continuous spectrum. 
Its structure can be understood from the dispersion equation 
$$
\omega=(1/2)k^2-\alpha_3 k^3
\eqno(81a)
$$
and the correspondig expression for the group velocity of continuous 
spectrum
$$
V_g(k)=k-3\alpha_3 k^2 \;.
\eqno(81b)
$$
From this we see that, at $\alpha_3>0$, the continuos radiation on the 
right hand side of the pulse is composed from 
$$
0<k\le k_0=1/6\alpha_3 \;,
\eqno(81c)
$$
where $V_g(k_0)=\text{max}V_g(k)$. As far as $k_0$ is less 
then the wavenumber of resonant radiation, 
approximately given by Eq.(50) (note that in our case , according to (45),
$a_2>1$), 
one can see why the peak of continuous spectra is less than 
$\kappa$ of resonant 
radiation. All this shows that the pulse in Fig. 13 is nothing than RRS. The 
time behavior of its amplitude is shown in Fig. 15 ; qualitatively, it is 
similar to Fig. 3a.

It is reasonable to compare these results with the case $\alpha_1=-0.6$, 
$\alpha_3=0$
, at the same initial condition. Then we have two regular solitons, 
propagating to the right. The soliton positions versus time for both cases 
are shown in Fig. 16. Note a very small acceleration of RRS in the direction 
of group velocity (which is now negative) . 
\begin{figure}[h] \vspace{-.5cm} \hspace{0.0cm}
\vspace*{3cm}
\caption{Soliton amplitude versus time at $\alpha_1=-0.6, \alpha_3=0.1$.}
\end{figure}

\bigskip

{\bf VII. CONCLUSIONS}

\bigskip

We have considered soliton and quasisoliton solutions of Eq. (1) and their 
relationship. Solutions (2), describing regular solitons, degenerate at 
$\alpha_3\to 0$; 
this shows that they exist due a balance between nonlinear terms and linear 
third order dispersion. On the other hand the quasisolitons (embedded and 
resonantly radiating) 
\begin{figure}[h] \vspace{-.5cm} \hspace{0.0cm}
\vspace*{3cm}
\caption{ Soliton positions versus time for $\alpha_1=-0.6, \alpha_3=0$ %
(1,2) and, $\alpha_1=-0.6, \alpha_3=0.1$ (3).}
\end{figure}
turn at $\alpha_3\to 0$ into the regular soliton solutions of Eq. (1) 
without the third derivative term. Apart from that, at $\alpha_2=0$, 
the regular solitons exist only in the 
special case $\alpha_1=6\alpha_3$. 

The resonantly radiating solitons (RRS) are nonsteady; the amplitudes of 
their radiation at small $\alpha_3$ are exponentially small and so their parameters 
change logarithmically slow in this case. Only such quasisteady solitons 
have sufficiently large lifetimes to be of physical significance. The 
embedded solitons (ES), which are steady structures consisting of pulses 
embedded into plane waves (wings), have close connection with RRS, similar 
to other systems. The corresponding analytical treatment in Sec.III is 
suplemented by numerical simulations with the cut off operation (Sec.IV) ; 
it is demonstrated that it leads to the transformation of ES into RRS that 
radiates in the direction of the radiation group velocity . Just after the 
cutt off, the amplitude, velocity and wavenumber of the radiated wavetrain 
coincide with those of wings; marching in time, we have seen that the 
amplitude decreases, the soliton velocity increases and the wavenumber 
changes according to Eqs.(30) and (31). This is in agreement with 
conservation laws (Appendix B). 

In Sec. V we have investigated a special case $\alpha_1=6\alpha_3$, 
interesting from a theoretical 
point of view . Then the quasisolitons, ES and RRS, do not exist and Eq. (1) 
can be reduced to the complex MKdV equation which has {\it N} -soliton 
solutions (consisting of regular solitons). This generalizes the Hirota 
solutions obtained for $\alpha_1=6\alpha_3$, $\alpha_2=0$.
The complex MKdV equation possesses the Painlev\'e property only at 
$\alpha_2=3\alpha_3$, which is the case when Eq.(1) is integrable.

Presumably, in the nonlinear {\it processes} only regular solitons and RRS 
may be of the physical significance. From Eq. (2b) it follows, however, that 
the amplitude of regular soliton vanishes at $\alpha_3 \to 0$, 
while for the RRS it remains 
finite. Therefore one should expect that at small $\alpha_3$, the RRS are more 
important in the nonlinear evolution. This was confirmed by solving the 
initial value problem (Sec. VI), where the RRS and not regular solitons are 
seen emerging from the initial disturbances. (Small $\alpha_3$ is the most interesting 
case from physical point of view because the third derivative term is in 
fact the result of an expansion) .

This work was partly supported from INTAS grant No 99-1068.

\bigskip

{\bf Appendix A}

\bigskip

{\bf Modulational instability of plane wave, according to Eq.(1)}

\bigskip

Eq. (1) has exact plain wave solution
$$
\Psi=A \text{exp}(i\kappa X - i\omega T) \;.
\eqno(A1)
$$
Substituting (A1) into (1), we have 
$$
\omega=\frac12\kappa^2-(1-\alpha_1\kappa)A^2-\alpha_3\kappa^3 \;.
\eqno(A2)
$$
Substituting a slightly perturbed wave (A1)
$$
\Psi=A(1-\chi)\text{exp}(i\kappa X-i\omega T)\;,
\eqno(A3)
$$
into Eq.(1), we have 
$$
i\chi_T+\frac12(1-6\alpha_3 K)\chi_{XX}+i(K+\alpha_1 A^2-3\alpha_3 K^2)\chi_X
$$
$$
+(1-\alpha_1\kappa)(\chi+\chi^*)A^2+i\alpha_2(\chi_X+\chi_X^*)A^2 \;.
\eqno(A4)
$$
Writing $\chi=u+iw$  and assuming that
$$
(u,w)\approx \text{exp}(ipX-irT)\;.
$$
we obtain dispersion equation, which is convenient to write in the form
$$
\Gamma^2-2\alpha_2kA^2\Gamma-\frac14(1-6\alpha_3K)k^2
[(1-6\alpha_3 K) k^2
$$
$$
-4(1-\alpha_1K)A^2]=0 \;,
\eqno(A5)
$$
where 
$$
\Gamma=r+\alpha_3p^3-(\kappa-3\alpha_3\kappa^2+\alpha_1A^2)p \;.
\eqno(A6)
$$
Thus, at real {\it p} , $\text{Im}\gamma=\text{Im}r$ and the stability condition is 
$$
(1-6\alpha_3\kappa)^2 p^2 \ge 4A^2[(1-\alpha_1\kappa)(1-6\alpha_3 \kappa)
-\alpha_2^2 A^2]\;. 
\eqno(A7)
$$
Unlike the plain wave solution of the regular NLS, now the plain wave can be 
stable at any {\it p} , if
$$
(1-\alpha_1\kappa)(1-6\alpha_3\kappa)\le \alpha_2^2 A^2 \;.
\eqno(A8)
$$
At $\alpha_2=0$ this is possible at
$$
\frac{1}{\alpha_1}\le \kappa \le \frac{1}{6\alpha_3}
\;\;\;\;\; (6\alpha_3<\alpha_1)
\eqno(A9a)
$$
$$
\frac{1}{6\alpha_3}\le \kappa \le \frac{1}{\alpha_1}
\;\;\;\;\; (\alpha_1<6\alpha_3)
\eqno(A9b)
$$
Applying the stability criterion to the wings of embedded solitons, where
$\kappa\approx 1/2\alpha_3$, 
we see that this may satisfy condition (A9b) ; then one can expect that the 
wings are stable. We should also take into account only $p>L^{-1}$, where {\it L} is 
the period in the numerical scheme; this relaxes the limitatations following 
from the stability criterion.

\bigskip

{\bf Appendix B}

\bigskip

{\bf Investigation of the soliton evolution by means of the integrals of 
motion.}

\bigskip

First , we present a general analysis of conserving integrals {\it N, P, H}. 
Substituting (5) into expressions (12), (14), (15) we have after simple 
algebra
$$
N=\int_{\infty}^{\infty} |\psi(x,t)|^2 dx \;, 
\eqno(B1)
$$
$$
P=KN+\frac12 i \int_{\infty}^{\infty}(\psi\partial_x \psi^*-
\psi^*\partial_x\psi) dx \;. 
\eqno(B2)
$$
where $\psi(x,t)$ satisfies Eq.(9). 

To have a convenient expression for {\it H}, we substitute in Eq. (15) 
$\alpha_3\partial_x^3\Psi$ and $\alpha_3\partial_x^3\Psi^*$ 
from Eq.(1), to obtain
$$
H=\frac{i}{2}\int_{\infty}^{\infty}dX(\Psi^*\partial_T\Psi-
\Psi\partial_T\Psi^*)+\frac12\int_{\infty}^{\infty}dX|\Psi|^4
$$
$$
+\frac{i}{4}\alpha_1\int_{\infty}^{\infty}dX|\Psi|^2
(\Psi^*\partial_X\Psi-\Psi\partial_X\Psi^*) \;.
$$
Then using the Galilei transformation in the form (24) and taking into 
account that at large $t$
$$
\partial_t\approx i(\lambda^2/2)\psi \;,
$$
we arrive at the following asymptotic expression
$$
H\approx\left(\Omega-KV-\frac12 \lambda^2  \right)N+PV+
\frac12 q \int_{\infty}^{\infty}|\psi|^4 dx 
$$
$$
-\frac12 i \alpha_1\int_{\infty}^{\infty}|\psi|^2
(\psi\partial_x \psi^*-\psi^*\partial_x\psi) dx \;. 
\eqno(B3) 
$$
Substituting (25) and (32) into (B1), we have at large {\it t} (cf. Refs. 
[25,26]),
$$
N\approx N_s+N_r \;, 
\eqno(B4)
$$
where 
$$
N_s=\int_{\infty}^{\infty}|u_s(x,t)|^2 dx \;,  \;\;\; 
N_r=\int_{\infty}^{\infty}|\eta(x,t)|^2 dx 
\eqno(B5)
$$
are contributions from the soliton and the radiation. Using (35) we have 
$$
N_s=\frac{2a_2}{\alpha_1}\text{arctan}\left(\frac{\alpha_1\lambda}
{\sqrt{a_2}q}\right)
\eqno(B6)
$$
and, at large {\it t},
$$
N_r \approx |f|^2|U|_t
\eqno(B7)
$$
where {\it f} is given by (52) or (55). Evidently, $|f|^2$ does not depend on 
{\it x}. Then from the conservation of {\it N} it follows
$$
\frac{d}{dt}\left[\frac{2a_2}{\alpha_1}\text{arctan}
\left(\frac{\alpha_1\lambda}{\sqrt{a_2}q}\right) \right]
\approx -|f|^2 |U| \;. 
\eqno(B8)
$$
In a similar way, from (B2) and the conservation of {\it P} we have the 
following asymptotic equation at large {\it t} 
$$
N\frac{dK}{dt}+\frac{d}{dt}\int_{\infty}^{\infty}u_s^2(x)\partial_x
\psi_s(x)dx +k|f|^2|U|\approx 0 \;.
\eqno(B9)
$$
From (36) at $\alpha_2=0$, we have 
$$
\partial_x\psi_s(x)=-\frac{\alpha_1}{2a_2} u_s^2(x) \;. 
\eqno(B10)
$$
Taking into account that
$$
\int_{\infty}^{\infty}u_s^4(x)dx=\frac{2a_2q}{\alpha_1}
\left(\frac{2\sqrt{a_2}\lambda}{q}-N_s\right),
\eqno(B11) 
$$
we transform Eq. (B9) to the form
$$
N\frac{dK}{dt}-\frac{d}{dt}\left[\frac{q}{\alpha_1}\left(
\frac{2\sqrt{a_2}\lambda}{q}-N_s\right)\right]
+k|f|^2|U|\approx 0 \;.
\eqno(B12)
$$

Now we turn to Eq.(B3). After differentiation over {\it t} and simple 
calculations we obtain
$$
P\frac{dV}{dt}-\left(\lambda\frac{d\lambda}{dt}+a_2K\frac{dK}{dt}\right)N
$$
$$
=-\frac{d}{dt}\left[
\frac12 q \int_{\infty}^{\infty}|\psi|^4 dx+
\frac{i}{4} \alpha_1 \int_{\infty}^{\infty}|\psi|^2
(\psi^*\partial_x \psi-\psi\partial_x\psi^*) dx
\right]
$$
$$
=-\frac{d}{dt}[
\frac12 q \int_{\infty}^{\infty}|\psi_s|^4 dx+
\frac{i}{4} \alpha_1 
$$
$$
\int_{\infty}^{\infty}|\psi_s|^2
(\psi_s^*\partial_x \psi_s-\psi_s\partial_x\psi_s^*) dx
]
\;. 
\eqno(B13)
$$
Using 
$$
\frac{dV}{dt}=a_2\frac{dK}{dt} \;, 
\eqno(B14)
$$
we obtain
$$
a_2(P-KN)\frac{dK}{dt}-\lambda\frac{d\lambda}{dt}N
\approx 
$$
$$
-\frac{d}{dt}\left(\frac12 q \int_{\infty}^{\infty}
u_s^4 dx + \frac{\alpha_1^2}{2a_2}
\int_{\infty}^{\infty} u_s^6 dx 
\right) \;. 
\eqno(B15)
$$
Let us apply these relations to small $\alpha_1$ 
(which is just the case shown in Figs.3). Assuming that
$$
\frac{\alpha_1 \lambda}{\sqrt{a_2}q}\ll 1 \;, 
\eqno(B16)
$$
we have 
$$
N_s\approx \frac{2\sqrt{a_2}\lambda}{q}
\left(1-\frac{\alpha_1^2\lambda^2}{3 a_2 q^2}
\right)\;. 
\eqno(B17)
$$
Therefore Eq. (B8) is reduced to
$$
\frac{d}{dt}\left[\frac{2\sqrt{a_2}\lambda}{q}
\left(1-\frac{\alpha_1^2\lambda^2}{3 a_2 q^2}\right)
\right]=-|f|^2|U| \;. 
\eqno(B18)
$$
In the same approximation, Eq.(12) takes the form 
$$
N\frac{dK}{dt}\approx-k|f|^2|U|+\frac{2\alpha_1}{3}
\frac{d}{dt}\left(\frac{\lambda^3}{\sqrt{a_2}q^2}
\right)\;, 
\eqno(B19)
$$
At $\alpha_1 \to 0$, Eqs. (B18) and (B19) turn into equations obtained in Ref.[18]. 
From (B19), (B14) and Eq. (60) it follows an important qualitative result (64).

Now, consider Eq. (B15). Taking into account (B2),(B10), (B11) and (B19) we 
see that at 
$$
t\ll\lambda k^{-2}|f|^{-4} U^{-2}
\eqno(B20)
$$
and small  $\alpha_1$, Eq. (B15) is approximately equivalent to (B18). Performing 
calculations similar to those in Ref. [19] and using (B14), we conclude 
that the soliton width $\lambda^{-1}$ and velocity $V(t)$
are changing logarithmically slow. The 
same holds for the soliton amplitude $u_(9t)$, determined from (39) and (37), in 
agreement with numerical solution of Eq.(1) (see Figs.3). The chracteristic 
time of the variations of soliton parameters has the same order of magnitude 
as in Ref.[26], which is much less than the right hand side of (B20). 

If $\alpha_1$ is not small, the analysis is rather tedious and will not be 
considered here.

\bigskip

{\bf References}

\bigskip

[1] A. Hasegawa and Y. Kodama, {\it Solitons in Optical Communications} 
(Oxford University Press, 1995). 

[2] G.P. Agrawal, {\it Nonlinear Fiber Optics}  (Academic Press, New York, 
1995).

[3] M.J. Potasek and M.Tabor, Phys.Lett. A {\bf 154} , 449 (1991). 

[4] R.Hirota, J. Math.Phys. {\bf 14} ,805 (1973).

[5] N.Sasa and J.Satsuma, J. Phys.Soc. Jpn. {\bf 60} ,409 (1991).

[6] E.M. Gromov, L.V. Piskunova and V.V.Tutin , Phys. Lett. A {\bf 256} , 
153 (1999) .

[7] J.Yang, B.A.Malomed and D.J. Kaup, Phys. Rev. Lett. {\bf 83}, 1958 
(1999).

[8] P.K.A.Wai, C.R.Menyuk, Y.C.Lee and Chen, Opt.Lett.11, 464 (1986).

[9] P.K.A.Wai, H.H.Chen and Y.C.Lee, Phys.Rev. A 41, 426 (1990).

[10] H.H. Kuehl and C.Y. Zhang, Phys.Fluids B 2, 889 (1990).

[11] V.I. Karpman, Phys. Rev.E , 47, 2073 (1993) .

[12] J.P. Boyd,{\it Weakly Nonlocal Solitary Waves and Beyond-All\_Order 
Asymptotics} (Kluver, Dodrecht,1998).

[13] V.I. Karpman, In : {\it Nonlinearity, Integrability and all 
that:Twenty} {\it years after NEEDS'79.} Editors M.Boiti, L.Martina, 
F.Pempinelli, B. Prinari \& G. Soliani, ("World Scientific", 2000), p.396.

[14] V.I. Karpman, Phys.Rev. E {\bf 62} , 5678 (2000).

[15] P.Deeskow, H.Schamel, N.N.Rao, M.Y. Yu, R.K. Varma and P.K.Shukla, 
Phys. Fluids {\bf 30} , 2703 (1987).

[16] V.I. Karpman and H. Schamel, Phys. Plasmas {\bf 4},120 (1997); 
ibid.{\bf 4} (1997) 2778 (E) .

[17] Y. Pomeau, A.Ramani and B. Grammaticos, Physica D {\bf 31} (1988) 127 .

[18] V.I.Karpman, Phys. Lett.A {\bf 181} (1993) 211.

[19] V.I. Karpman, Phys.Lett.A {\bf 186} (1994) 300.

[20] V.I.Karpman, Phys.Rev.Lett. {\bf 74} (1995) 2455.

[21] J. Weiss, M.Tabnd G. Carnevale, J. Math. Phys. {\bf 24} , 522 (1983) .

[22] M.A. Ablowitz and H. Segur {\it Solitons and the Inverse Scattering 
Transform} (SIAM, Philadelphia, 1981)

[23] M. Gedalin, T.C. Scott and Y.B. Band, Phys. Rev. Lett. {\bf 78}, 448 
(1997). 

[24] V. E. Zakharov and E.A. Kuznetsov, JETP {\bf 86},1036 (1998).

[25] V.I. Karpman and A.G. Shagalov, Phys. Lett. A {\bf 254} , 319 (1999).

[26] V.I.Karpman, Phys. Lett. A {\bf 244} , 397 (1998) .{\bf} 

[27] Zhonghao Li, Lu Li, Huipping Tian and Guosheng Zhou, Phys.Rev. 

Lett. {\bf 84} ,4096 (2000). 

[28] K. Portezian and K. Nakkeeran, Phys.Rev.Lett. {\bf 76}, 3955 (1996)

\end{document}